\documentclass[]{elsart}
\usepackage{graphicx,amssymb,lineno}
\usepackage{bm}
\usepackage{epsfig}
\usepackage{color}

\newcommand{\joerg}[1]{\textcolor{black}{#1}}

\begin{document}
\begin{frontmatter}
\title{Statistics of the Energy Dissipation Rate and Local Enstrophy in Turbulent Channel Flow}
\author{Peter E. Hamlington}
\address{Laboratory for Computational Physics and Fluid Dynamics, Naval Research Laboratory, Washington D.C.\  20375, USA}

\author{Dmitry Krasnov, Thomas Boeck, J\"org Schumacher}
\address{Institute of Thermodynamics and Fluid Dynamics,
Ilmenau University of Technology, P.O.\ Box 100565 D-98684
Ilmenau, Germany}

\begin{abstract}
Using high-resolution direct numerical simulations, the height and Reynolds number dependence of higher-order statistics of the energy dissipation rate and local enstrophy are examined in incompressible, fully-developed turbulent channel flow. The statistics are studied over a range of wall distances, spanning the viscous sublayer to the channel flow centerline, for friction Reynolds numbers $Re_\tau\!=\!180$ and $Re_\tau\!=\!381$. The high resolution of the simulations allows dissipation and enstrophy moments up to fourth order to be calculated. These moments show a dependence on wall distance, and Reynolds number effects are observed at the edge of the logarithmic layer. Conditional analyses based on locations of intense rotation are also carried out in order to determine the contribution of vortical structures to the dissipation and enstrophy moments. Our analysis shows that, for the simulation at the larger Reynolds number, small-scale fluctuations of both dissipation and enstrophy become relatively constant for $z^+\!\gtrsim \!100$. 
\end{abstract}

\end{frontmatter}

\section{Introduction}
\label{intro}

Determining the detailed properties of wall-bounded turbulent shear flows has been the focus of considerable experimental and computational research (see, e.g., \cite{panton2001,adrian2007,wallace2009} for reviews). Although prior studies have provided insights into low-order statistics and coherent structures in these flows, higher-order velocity gradient statistics remain relatively unexplored. Velocity gradients reflect the structure and properties of the turbulent small scales, \cite{tsinober2009} and their higher-order statistics are particularly sensitive to the large-amplitude, intermittent fluctuations characteristic of high-Reynolds number flows. While such higher-order statistics have been studied in homogeneous isotropic turbulence using direct numerical simulations (DNS) for several decades, \cite{vincent1991,jimenez1993,sreenivasan1997,donzis2008,ishihara2009} it has only recently become feasible to carry out similar analyses in wall-bounded flows. This is due, in large part, to the substantial computational resources required to resolve the smallest scales of the turbulence. Progress in simulating wall bounded flows has recently been made, however, in Ref.\ \cite{boeck2010}, where velocity gradient moments up to fourth order are examined using DNS of fully-developed turbulent channel flow at $Re_\tau\!=\!180$, where $Re_\tau\!=\!u_\tau L/\nu$, $u_\tau$ is the friction velocity, $L$ is the half-width of the channel, and $\nu$ is the kinematic viscosity.

The present paper refines and significantly extends the prior study in Ref.\ \cite{boeck2010} by using highly resolved turbulent channel flow DNS at $Re_\tau\!=\!180$ and $Re_\tau\!=\!381$ to examine both the height and Reynolds number dependence of higher-order velocity gradient statistics. Particular emphasis is placed on the moments of the energy dissipation rate 
\begin{equation}\label{diss}
\varepsilon=2\nu S'_{ij} S'_{ij}\,,
\end{equation} 
and the local enstrophy (or enstrophy density) 
\begin{equation}\label{enst}
\Omega = \frac{1}{2} \omega'_i \omega'_i\,.
\end{equation}
Taken together, these quantities characterize the straining and rotation associated with small-scale turbulent fluctuations. \cite{donzis2008} The fluctuating strain rate tensor, $S'_{ij}$, in Eq.\ (\ref{diss}) is given in terms of the fluctuating velocity, $u'_i\!=\!u_i - \langle u_i\rangle$ (where $\langle \cdot \rangle$ is a $z$-dependent average over time and $x$-$y$ planes parallel to the channel walls), as
\begin{equation}\label{sijf}
S'_{ij} = \frac{1}{2} \left(\frac{\partial u'_i}{\partial x_j} +
\frac{\partial u'_j}{\partial x_i}\right)\,,
\end{equation}
and the fluctuating vorticity, $\omega'_i$, in Eq.\ (\ref{enst}) is given by
\begin{equation}
\omega'_i = \epsilon_{ijk}\frac{\partial u'_k}{\partial x_j}\,,
\end{equation} 
where $\epsilon_{ijk}$ is the cyclic permutation tensor. The moments of $\varepsilon$ and $\Omega$, denoted $\langle \varepsilon^n\rangle$ and $\langle \Omega^n \rangle$, reflect properties of small-scale, high-amplitude fluctuations in the velocity gradient. For example, $\langle \varepsilon^4 \rangle/\langle \varepsilon^2 \rangle^2$ and $\langle \Omega^4 \rangle/\langle \Omega^2 \rangle^2$ can be used to quantitatively assess the degree of small-scale intermittency in the flow. \cite{tsinober2009} 

In order to measure moments of $\varepsilon$ and $\Omega$ up to fourth order, the resolutions used in the present simulations are substantially higher than in prior simulations at similar values of $Re_\tau$. Recent studies of homogeneous isotropic turbulence \cite{donzis2008,schumacher2007} and turbulent channel flow \cite{boeck2010} have shown that very fine resolutions are required to fully resolve higher-order statistics of $\varepsilon$ and $\Omega$. In particular, using simulations of a $Re_\tau\!=\!180$ channel flow at three different resolutions, it was shown in Ref.\ \cite{boeck2010} that the lowest resolution was sufficient to measure low-order statistics such as mean velocities and Reynolds stresses, $\overline{u'_i u'_j}$. Finer resolutions, however, were required to accurately measure higher-order velocity gradient statistics such as the mean value of $\varepsilon$, particularly near the channel walls. The resolutions used in the present simulations are even higher than those in Ref.\ \cite{boeck2010}, and allow calculation of up to fourth order moments of $\varepsilon$ and $\Omega$.

A wide range of wall distances are examined in the present study in order to determine the height dependence of the statistics. Turbulent channel flows are typically divided into four regions; the viscous sublayer for $z^+\!<\!5$, the buffer region for $5\!<\!z^+\!<\!30$, the logarithmic layer for $z^+\!>\!30$ and $z\!\lesssim\!0.3 L$, and the outer flow for $z\!\gtrsim\!0.3 L$, \cite{pope2000} where $z^+\!\equiv\! z u_\tau/\nu$ and $z$ is the coordinate in the wall-normal direction. For both values of $Re_\tau$ considered here, statistics are examined from $z^+\!=\!2$, well within the viscous sublayer, to the channel centerline. The inhomogeneity in the $z$-direction requires that the statistics of $\varepsilon$ and $\Omega$ be calculated in planes parallel to the channel walls, in order to allow averaging over homogeneous flow directions. Compared to studies of three-dimensional homogeneous isotropic turbulence, where full volume averaging is possible, this places significant restrictions on the statistical convergence of the results. Consequently, particular attention is paid in the following to the convergence of the moments of $\varepsilon$ and $\Omega$ as a function of $z^+$ and $Re_\tau$. 

While variations in the moments of $\varepsilon$ and $\Omega$ with $Re_\tau$ and $z^+$ are important for understanding the small-scale structure of the channel flow, the presence of coherent vortical structures is also expected to play a role in determining these moments. Theodorsen \cite{theodorsen1952} first proposed a hairpin shape for these structures, and subsequent experimental (e.g., Refs.\ \cite{head1981,ong1998,ganapathi2006}) and numerical studies (e.g., Refs.\ \cite{kim1987,chong1998,pirozzoli2008}) have characterized hairpin vortices throughout wall bounded flows (see Ref.\ \cite{adrian2007} for a review). Although these vortices have been identified as important in fluid transport and the generation of Reynolds stresses, \cite{pirozzoli2008} relatively little is known about their contribution to the higher-order, small-scale statistics of $\varepsilon$ and $\Omega$. In the following, we examine this issue using conditional analyses of the moments of $\varepsilon$ and $\Omega$ based on locations of intense rotation in the flow. We also consider the moments at locations away from regions of intense rotation in order to determine the extent to which differences in the moments as a function of $Re_\tau$ and $z^+$ can be attributed to vortical structures.

The manuscript is organized as follows. Details of the numerical simulations are presented in the next section. The moments of $\varepsilon$ and $\Omega$ are then presented up to fourth order for both values of $Re_\tau$, accompanied by an analysis of the statistical convergence of these moments. Conditional analyses based on locations of intense rotation are outlined in Section \ref{struct}. The method by which these locations are identified is briefly discussed, and results from conditional analyses of the $\varepsilon$ and $\Omega$ moments are presented. Finally, a summary and conclusions are provided at the end.

\section{Description of Numerical Simulations}
\label{nsims}

The numerical simulations used in the present study solve the incompressible Navier Stokes equations for a fully developed turbulent channel flow. These equations are written in non-dimensional form as
\begin{eqnarray}\label{ns1}
&&\frac{\partial u_i}{\partial x_i}=0\,,\\
&&\frac{\partial u_i}{\partial t} + u_j \frac{\partial u_i}{\partial x_j}
= - \frac{\partial p}{\partial x_i} +
\frac{1}{Re} \frac{\partial^2 u_i}{\partial x_j \partial x_j}\,, \label{ns2}
\end{eqnarray} 
where $u_i$ is the velocity field and $p$ is the kinematic pressure. The Reynolds number, $Re$, is given as $Re\!=\!U L/\nu$, where $U$ is the total mean velocity in the channel. 

As described in Ref.\ \cite{boeck2010}, the Navier-Stokes equations in Eqs.\ (\ref{ns1}) and (\ref{ns2}) are decomposed in poloidal-toroidal form and then solved using a pseudo-spectral method. This method uses Fourier expansions in the $x$ and $y$ directions, which are parallel to the channel walls, and Chebyshev polynomial expansions in the $z$ direction, normal to the walls. \cite{gottlieb1977,canuto1988} De-aliasing with the $2/3$ rule is applied in all three directions. \joerg{The time discretization combines an implicit backward-differentiation method for linear terms with an explicit Adams-Bashforth method for nonlinear terms, and is second-order accurate. The method can be symbolically written as
\begin{equation}\label{time}
\frac{3 f^{n+1}-4 f^n + f^{n-1}}{2 \Delta t}= {\cal L} f^{n+1} 
+ 2{\cal N}(f^n)-{\cal N}(f^{n-1})\,,
\end{equation}
where $\cal L$ is a linear operator, $\cal N$ represents nonlinear terms, and $\Delta t$ is the time step. The approximation for the time derivative, $\partial f/\partial t$, on the left side of Eq.\ (\ref{time}) gives $f$ at time level $n+1$ using values at the two previous levels, denoted $f^n$ and $f^{n-1}$. The simulations are parallelized using MPI.}

\begin{table}[t]
\begin{center}
\begin{tabular}{@{}cccccccc@{}}
$Re_{\tau}$ &  $Re$ & $N_x\!\times\! N_y\!\times\! N_z$ & $N_t$ & $\Delta x^+$ & $\Delta y^+$ & $\Delta z_c^+$\\
\hline
$ 180$ & $2800$ & $ 512\!\times\! 512\!\times\!1025$ &  $403$ & $ 4.4$ & $ 2.2$ & $0.55$\\
$ 381$ & $6667$ & $1024\!\times\!512\!\times\!1025$ & $134$ & $ 2.3$ & $ 2.3$ & $1.2$\\
\hline
\end{tabular}
\caption{Friction Reynolds number  $Re_{\tau}$, global Reynolds number $Re\!=\!U L/\nu$, grid dimensions $N_x\!\times\! N_y\!\times\! N_z$, number of temporal snapshots, $N_t$, horizontal resolutions, $\Delta x^+\! \equiv\! \Delta x u_\tau/\nu$ and $\Delta y^+\!\equiv\! \Delta y u_\tau/\nu$, and vertical resolution at the centerline, $\Delta z_c^+\! \equiv\! \Delta z(z/L\!=\!1) u_\tau/\nu$, for the two simulations analyzed herein.}
\label{resolution}
\end{center}
\end{table}

The channel flow simulation domain consists of a rectangular box of non-dimensional size $(L_x\! \times\! L_y\! \times\! L_z)/L\! =\! 4\pi\! \times\! 2\pi \!\times\! 2$ for $Re_\tau\!=\!180$ and $2 \pi\! \times\! \pi\! \times\! 2$ for $Re_\tau\!=\!381$. In Ref.\ \cite{boeck2010}, the $Re_\tau\!=\!180$ case was examined at a maximum resolution of $N_x\! \times\! N_y\! \times\! N_z\! =\! 512\! \times\! 512\! \times\! 513$ collocation points in physical space. Due to de-aliasing, the number of Fourier or Chebyshev modes is $2/3$ the grid numbers $N_{x,y,z}$. In the present study, this resolution has been increased to $512\! \times\! 512 \!\times\! 1025$ collocation points in order to achieve the best possible convergence of the velocity gradient statistics. The resolution in the $Re_\tau\!=\!381$ case, which was not examined in Ref.\ \cite{boeck2010}, is $1024\! \times\! 512 \!\times\! 1025$ collocation points (see Table \ref{resolution}). \joerg{These parameters give grid spacings that are approximately two to five times finer than in prior simulations at comparable values of $Re_\tau$. The $Re_\tau\!=\!180$ channel flow simulated by Kim \textit{et al.} \cite{kim1987}, for example, used $192\times 160\times 129$ grid points in a domain of size $4\pi\! \times\! 2\pi\! \times\! 2$ (where, consistent with the present study, $z$ is used as the wall-normal coordinate). Similarly, the $Re_\tau\!=\!395$ channel simulation by Moser \textit{et al.} \cite{moser1999}) was performed on a grid consisting of $256\!\times\! 192\!\times\! 193$ points in a domain of size $2\pi\! \times\! \pi\! \times\! 2$.} 

Temporal snapshots of the flow field are stored every $0.1$ convective time units, $L/U$, for the $Re_\tau\!=\!180$ simulation, with a total of $N_t\!=\!403$ snapshots. The time interval between snapshots for the $Re_\tau\!=\!381$ simulation is $0.19$ convective time units, with $N_t\!=\!134$ total snapshots. \joerg{The statistical analysis for a single horizontal plane (and its symmetry counterpart in the upper half of the channel) can thus be made over a set of up to $2.1\times 10^8$ data points. Due to the larger $N_t$ for the $Re_\tau\!=\!180$ case and the greater $N_x$ for $Re_\tau\!=\!381$ (see Table \ref{resolution}), we use a similar amount of data in the analysis of each value of $Re_\tau$.}

\begin{figure}[t]
\begin{center}
\includegraphics[width=6.3cm]{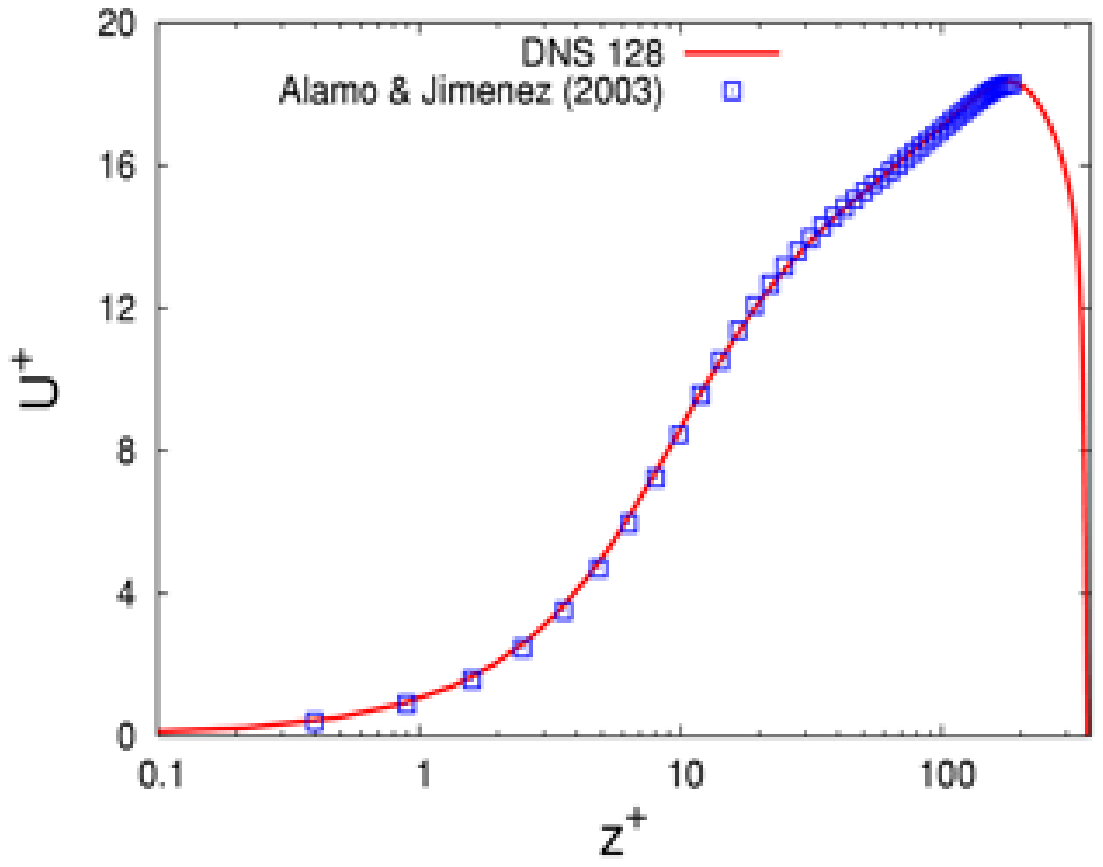}
\includegraphics[width=6.3cm]{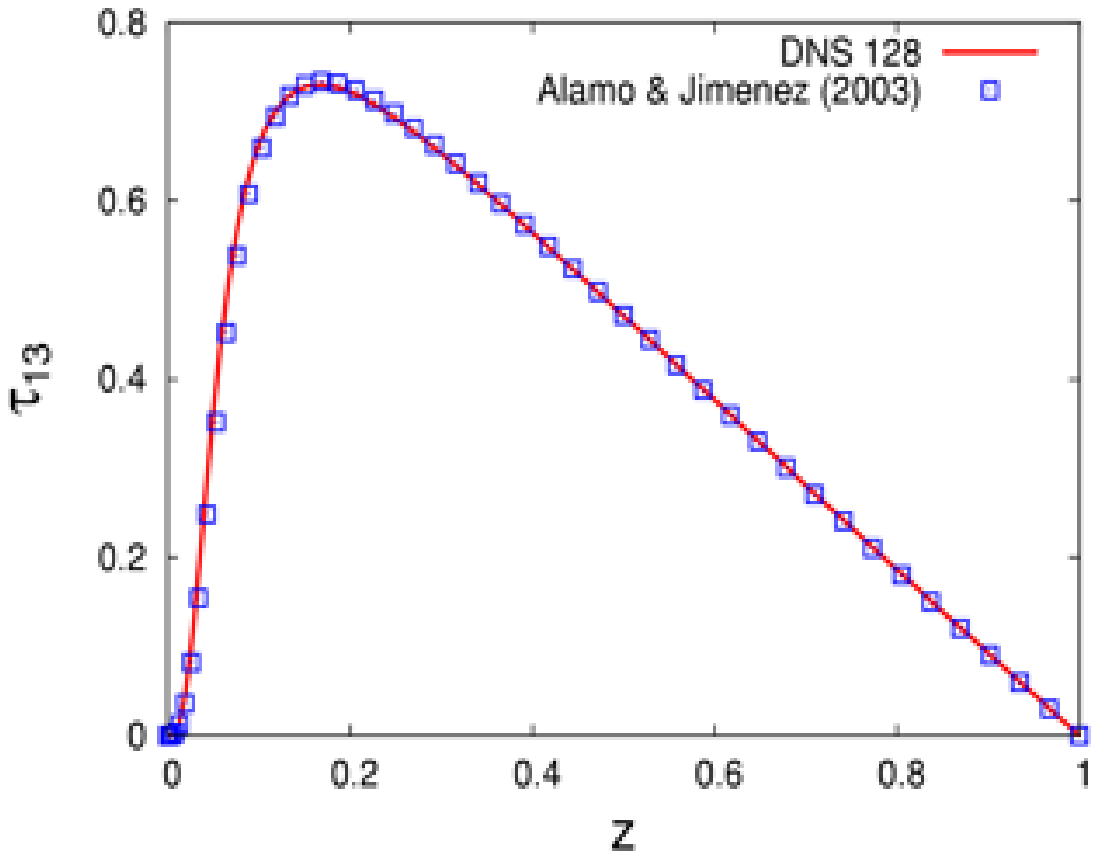}
\caption{Mean streamwise velocity $\overline{u}^+$ (a) and Reynolds shear stress $\tau^+_{13}\!=\!\overline{u'_x u'_z}^+$ (b) from $Re_\tau\!=\!180$ numerical simulation at resolution $128\!\times\! 128\!\times\! 129$. Results show good agreement with DNS of fully-developed turbulent channel flow by del Alamo and Jim\'{e}nez \cite{alamo2003} for $Re_\tau\!=\!180$.}\label{mean_vars}
\end{center}
\end{figure}

Figure \ref{mean_vars} shows the mean velocity in the $x$ direction and the Reynolds shear stress $\tau_{13}^+\!=\!\overline{u'_x u'_z}^+\!=\!\overline{u'_x u'_z}/u_\tau^2$ as a function of $z^+$ for $Re_\tau\!=\!180$ at a resolution of $128\!\times\! 128\!\times\! 129$. Even for this relatively low resolution, the simulation results are in good agreement with recent channel flow simulations by del Alamo and Jim\'{e}nez \cite{alamo2003}. The high resolution of the present simulations further allows the higher-order statistics of $\varepsilon$ and $\Omega$ to be calculated.

\section{Statistics of Energy Dissipation Rate and Local Enstrophy}
\label{stats}

The statistics of $\varepsilon$ and $\Omega$ are examined in the following at various wall distances, $z^+$, by carrying out the analysis in $x$-$y$ planes parallel to the channel walls. Within each of these planes, the flow is essentially homogeneous, thus allowing an examination of the flow statistics similar to that employed for homogeneous isotropic turbulence. In the following, the average $\langle \cdot \rangle$ denotes an $x$-$y$ average at a particular value of $z^+$. It is always combined with an arithmetic average over the full sequence of temporal snapshots. \joerg{Statistical convergence is further improved by using symmetric planes from both the top and bottom halves of the channel, where the velocity and velocity gradient fields in the top half are reflected about the centerline.}

 \begin{figure}[t]
\begin{center}
\includegraphics{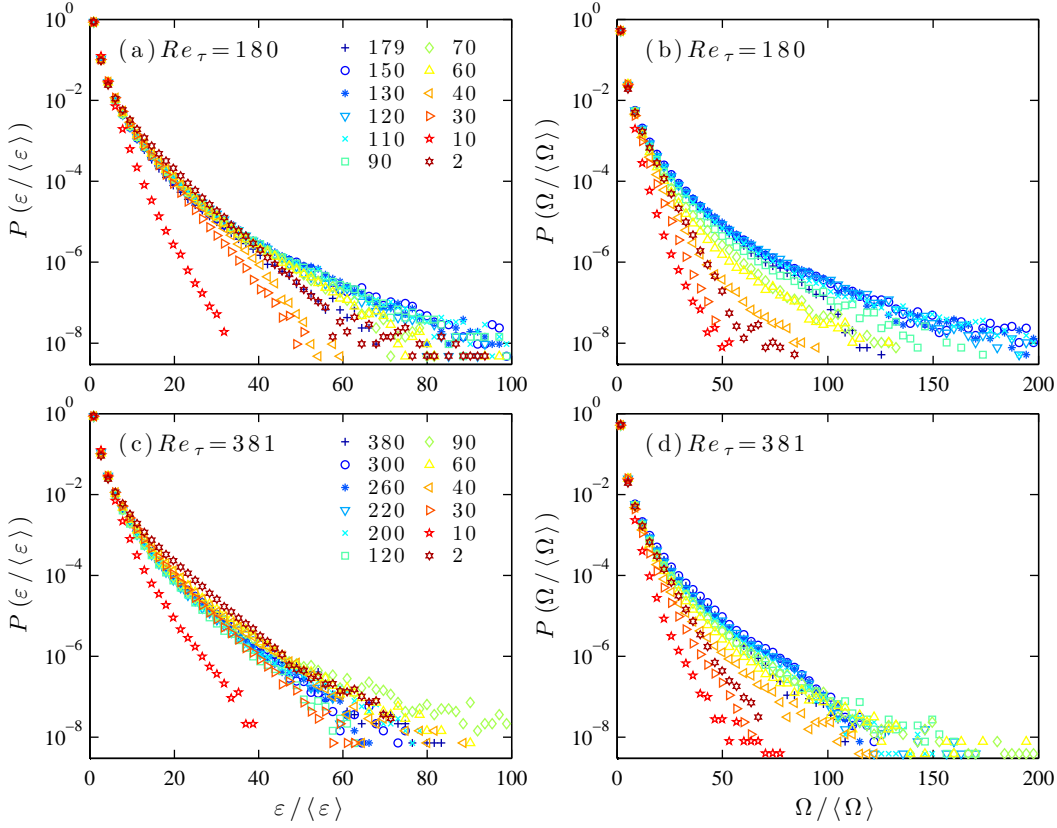}
\caption{Probability distribution functions of dissipation rate $\varepsilon/\langle \varepsilon\rangle$ and local enstrophy $\Omega/\langle \Omega\rangle$ for $Re_\tau\!=\!180$ ($\varepsilon$: (a) and $\Omega$: (b)) and $Re_\tau=381$ ($\varepsilon$: (c) and $\Omega$: (d)).}\label{de}
\end{center}
\end{figure}

\subsection{Distributions of the Dissipation and Enstropy}
\label{dist} 

Figure \ref{de} shows probability density functions (pdfs) of $\varepsilon/\langle \varepsilon \rangle$ and $\Omega/\langle \Omega\rangle$ for both values of $Re_\tau$ at wall distances spanning the viscous sublayer ($z^+\!=\!2$) to the channel centerline, where the averages $\langle \varepsilon \rangle$ and $\langle \Omega\rangle$ are functions of $z^+$. The pdfs of both $\varepsilon$ and $\Omega$ vary as the wall is approached from the centerline, and Figure \ref{de} shows that the degree of intermittency, as indicated by the width of the pdf tails and the corresponding probability of obtaining large amplitudes of $\varepsilon$ and $\Omega$, generally decreases as the wall is approached, with a minimum in the buffer layer at approximately $z^+\!=\!10$. Immediately at the wall, for $z^+\!=\!2$, however, the pdfs of $\varepsilon$ and $\Omega$ for both values of $Re_\tau$ are wider than those for $z^+\!=\!10$. For $Re_\tau\!=\!381$, the tails of the $\varepsilon$ and $\Omega$ pdfs are most prominent in the logarithmic layer for $z^+\!\approx\! 60-90$, while for $Re_\tau\!=\!180$ the most prominent tails occur further from the wall at $z^+\!\approx\!90-150$. This is, in turn, indicative of increased intermittency at these values of $z^+$, and may be connected to the bursting of coherent vortical structures away from the wall. \cite{adrian2007} 

With respect to the differences between the pdfs of $\varepsilon$ and $\Omega$, Figure \ref{de} shows that for both values of $Re_\tau$ \joerg{and wall distances down to $z^+\!\approx\!40$}, the tails of the $\Omega$ pdfs are more pronounced than those for the $\varepsilon$ pdfs. This is consistent with prior results in homogeneous isotropic turbulence \cite{donzis2008} and turbulent channel flow \cite{boeck2010}. \joerg{For $z^+\!\lesssim\!40$, however, the pdfs of $\varepsilon$ and $\Omega$ are more similar, indicating a closer correspondence between the statistics of $\varepsilon$ and $\Omega$ near the wall.}
 
\begin{figure}[t]
\begin{center}
\includegraphics[]{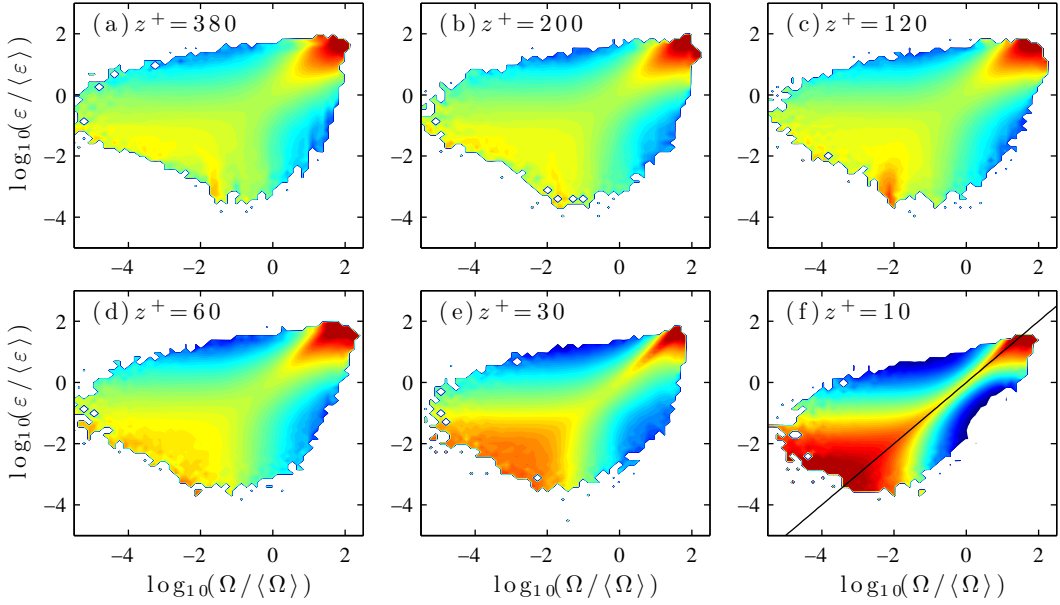}
\caption{Joint pdfs of $\varepsilon$ and $\Omega$ at six values of $z^+$ for $Re_\tau=381$. The joint pdfs are normalized by the one-dimensional pdfs of $\varepsilon$ and $\Omega$, as in Eq.\ (\ref{pidef}), and $\log_{10}[\Pi(\varepsilon,\Omega)]$ is shown. The color contours are the same in all panels, and range from $10^{-4}$ (blue) to $10^3$ (red). The black diagonal line in panel (f) corresponds to the relation $\varepsilon \!=\! 2\nu \Omega$ obtained from Eq.\ (\ref{epsom}).}\label{de_joint}
\end{center}
\end{figure}
 
The connection between $\varepsilon$ and $\Omega$ at each value of $z^+$ can be  further examined using joint probability distributions of $\varepsilon$ and $\Omega$, shown in Figure \ref{de_joint} for $Re_\tau\!=\!381$. Similar joint pdfs are shown for $Re_\tau\!=\!180$ in Ref.\ \cite{boeck2010}. The joint pdfs, denoted $P(\varepsilon,\Omega)$, are normalized by $P(\varepsilon)$ and $P(\Omega)$ as \cite{boeck2010}
\begin{equation}\label{pidef}
\Pi(\varepsilon,\Omega) = \frac{P(\varepsilon,\Omega)}{P(\varepsilon)P(\Omega)}\,,
\end{equation}
where values of $\Pi(\varepsilon,\Omega)$ greater than $1$ indicate a higher correlation between $\varepsilon$ and $\Omega$ than if the two quantities were statistically independent. \cite{zeff2003} Consistent with prior results for $Re_\tau\!=\!180$ \cite{boeck2010} and with simulations of homogeneous isotropic turbulence, \cite{zeff2003} Figure \ref{de_joint} shows that intense $\varepsilon$ and $\Omega$ are correlated at all values of $z^+$. As the wall is approached, however, Figure \ref{de_joint} shows that the support of the joint pdfs decreases. As noted in Ref.\ \cite{boeck2010}, the averages of $\varepsilon$ and $\Omega$ are connected in a wall-bounded shear flow by
\begin{equation}\label{epsom}
\langle \varepsilon\rangle = 2\nu \langle \Omega\rangle + 2\nu \frac{\partial^2 \langle u'^2_z\rangle}{\partial z^2}\,.
\end{equation}
Since the second term on the right-side of (\ref{epsom}) is small compared to $2\nu\langle \Omega\rangle$ in the channel flow, \cite{boeck2010} we obtain $\langle \varepsilon\rangle \!\approx\! 2\nu \langle \Omega\rangle$. Figure \ref{de_joint} shows that the values of $\varepsilon$ and $\Omega$ fall increasingly close to this relation as the wall is approached. These results at $Re_\tau=381$ are qualitatively consistent with the findings in Ref.\ \cite{boeck2010} for $Re_\tau=180$.

\subsection{Moments of the Dissipation and Enstrophy}
\label{de_moms}

\begin{figure}[t]
\begin{center}
\includegraphics{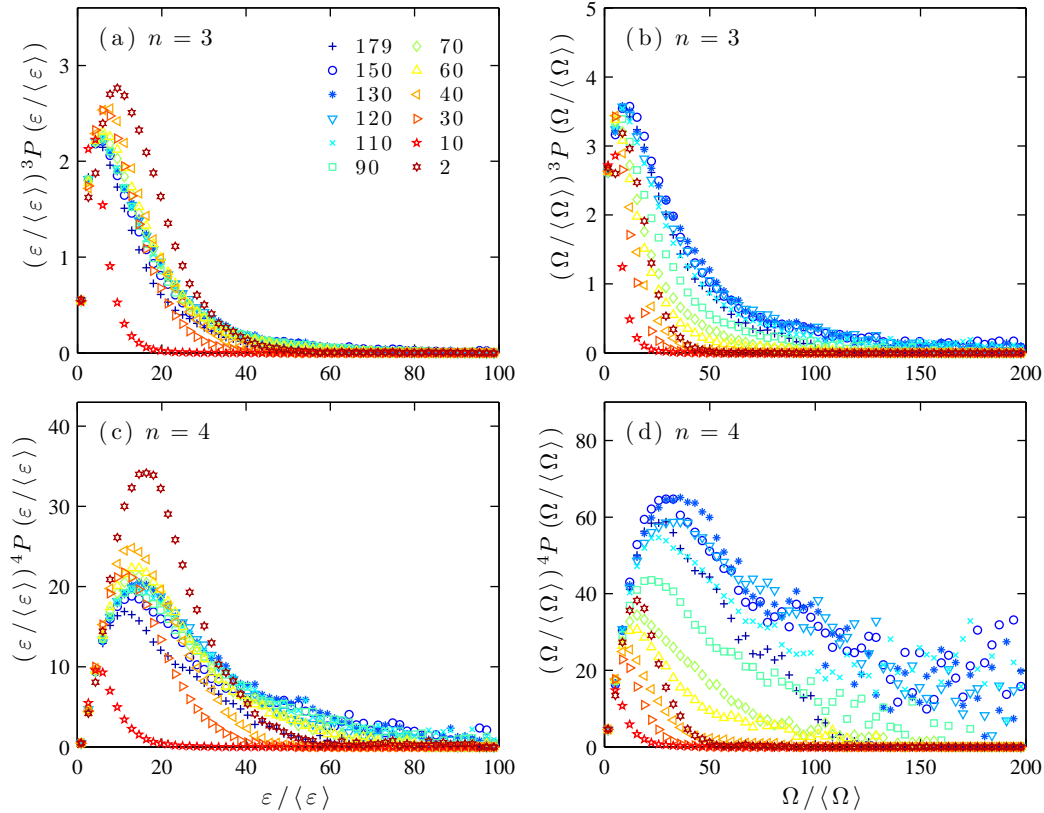}
\caption{Moment convergence results for $\varepsilon$ and $\Omega$ for $Re_\tau=180$. The $n\!=\!3$ and $n\!=\!4$ moments are shown for $\varepsilon$ (a) and (c), and $\Omega$ (b) and (d), respectively.}\label{de_conv180}
\end{center}
\end{figure}

\begin{figure}[t]
\begin{center}
\includegraphics{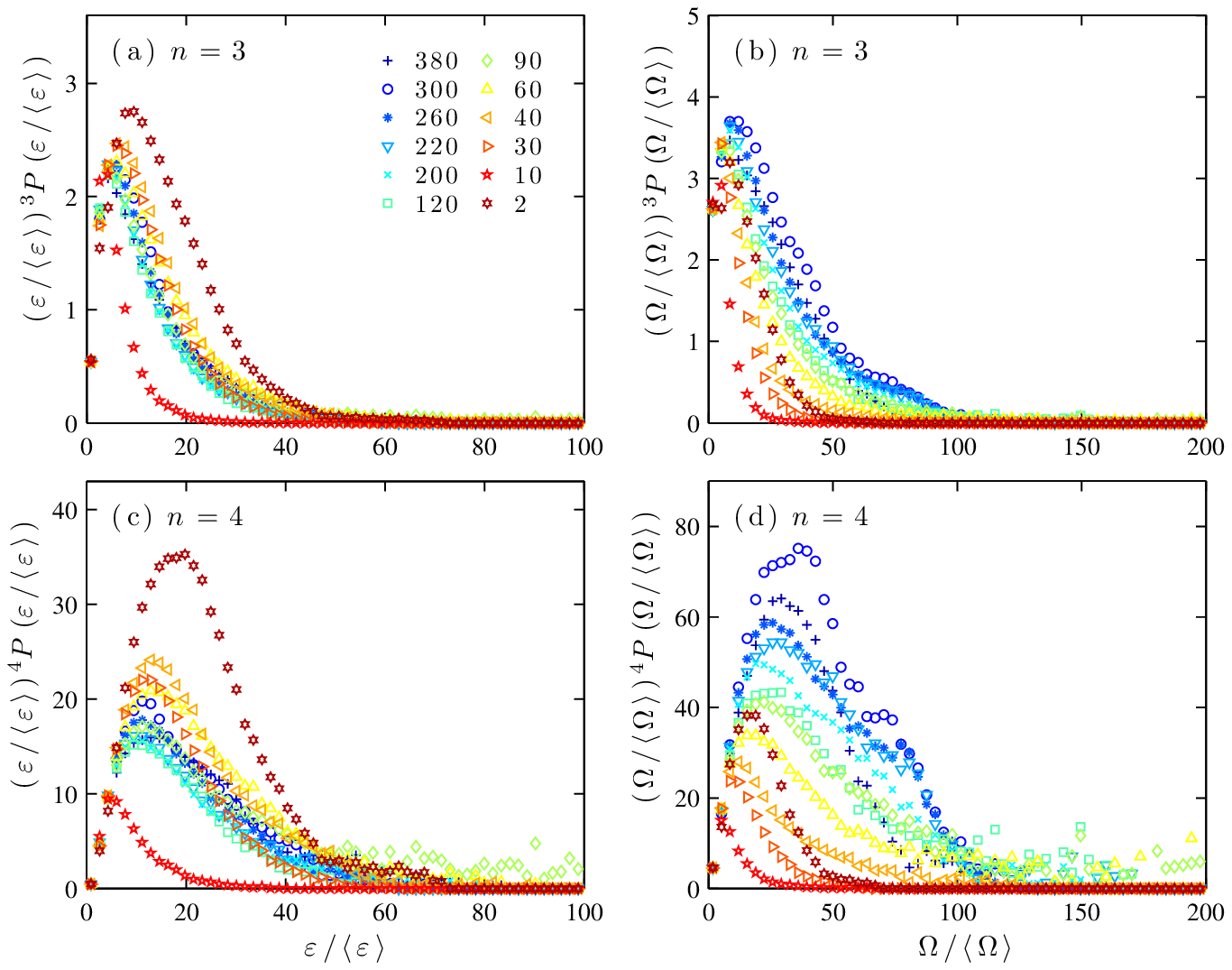}
\caption{Moment convergence results for $\varepsilon$ and $\Omega$ for $Re_\tau=381$. The $n\!=\!3$ and $n\!=\!4$ moments are shown for $\varepsilon$ (a) and (c), and $\Omega$ (b) and (d), respectively.}\label{de_conv}
\end{center}
\end{figure}

The statistical convergence of a moment $\langle \varphi^n\rangle$ can be assessed from plots of $\varphi^n P(\varphi)$ versus $\varphi$, where $\varphi$ is an arbitrary velocity gradient quantity such as $\varepsilon$ or $\Omega$. \cite{donzis2008} These distributions are shown for $n\!=\!3$ and $n\!=\!4$ for the $Re_\tau\!=\!180$ and $Re_\tau\!=\!381$ simulations in Figures \ref{de_conv180} and \ref{de_conv}, respectively. \joerg{Figure \ref{de_conv180} shows that, for $Re_\tau\!=\!180$, the $n\!=\!3$ moments are converged for both $\varepsilon$ and $\Omega$ at all $z^+$, as indicated by the decrease in the curves to zero for large $\varepsilon/\langle \varepsilon \rangle$ and $\Omega/\langle \Omega\rangle$. While the $n\!=\!4$ moments of $\varepsilon$ are also converged at all $z^+$, the $n\!=\!4$ moments of $\Omega$ show a lack of convergence for $z^+\!\approx\!90\!-\!150$. These wall distances correspond to the pdfs with the longest tails in Figure \ref{de}(b), and the lack of convergence shown in Figure \ref{de_conv180}(d) is likely due to the bursting of vortical structures across the channel. This would also explain why the higher-order $\Omega$ moments are more strongly affected than the $\varepsilon$ moments, and suggests that the non-converged fourth moments of $\Omega$ in Figure \ref{de_conv180}(d) may have a physical origin.} 

\begin{figure}[t]
\begin{center}
\includegraphics[]{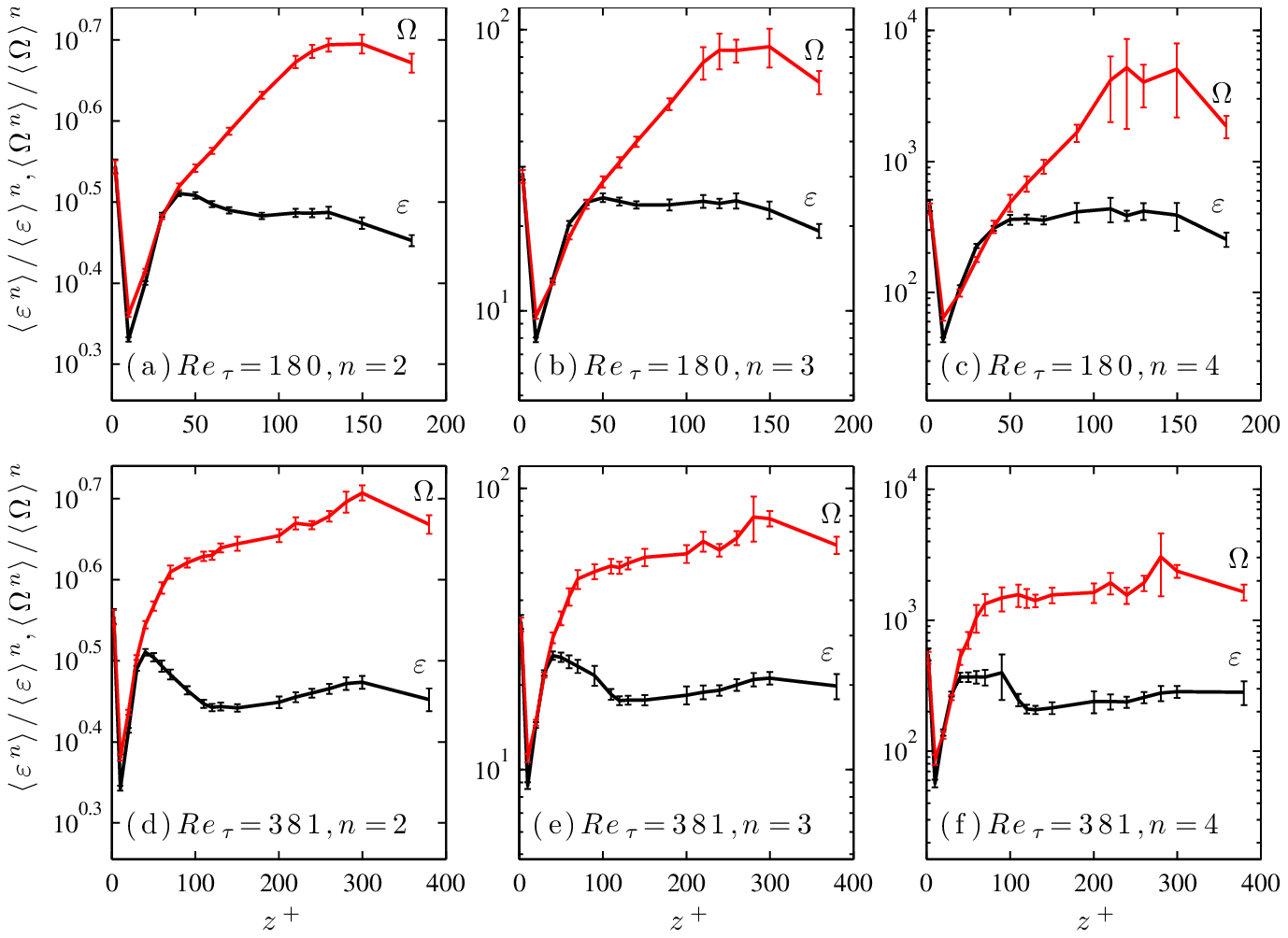}
\caption{Moments $\langle \varepsilon^n\rangle /\langle \varepsilon\rangle^n$ (black lines) and $\langle \Omega^n\rangle /\langle \Omega\rangle^n$ (red lines) as a function of $z^+$ for $n\!=\!2-4$ in the $Re_\tau\!=\!180$ (a)-(c) and $Re_\tau\!=\!381$ (d)-(f) simulations. Legend is shown in panel (c). \joerg{Error bars show estimates of 90\% confidence intervals obtained using a moving-block bootstrap method, \cite{garcia2005} with block sizes equal to one convective time unit ($L/U$) for both values of $Re_\tau$.}}\label{moms}
\end{center}
\end{figure}

For $Re_\tau\!=\!381$, Figures \ref{de_conv}(a) and (c) show that the moments of $\varepsilon$ are adequately resolved up to $n\!=\!4$ at all values of $z^+$. \joerg{Figures \ref{de_conv}(b) and (d) further show that the $n\!=\!3$ and $n\!=\!4$ moments of $\Omega$ are generally better converged than those for $Re_\tau\!=\!180$ in Figure \ref{de_conv180}. Given the fact that approximately the same amount of data is used in the analysis of both runs, this suggests that the cross-channel bursting of vortex structures may be less prevalent for higher $Re_\tau$. In particular, there is an extended bulk region for the higher Reynolds number case that separates the two logarithmic layers in the upper and lower halves of the channel.} 

Figure \ref{moms} shows the resulting moments for $n\!=\!2\!-\!4$ for the $Re_\tau\!=\!180$ and $Re_\tau\!=\!381$ simulations throughout the entire channel. \joerg{Estimates of 90\% confidence intervals for the calculated moments have been included in Figure \ref{moms} using a moving-block bootstrap method \cite{garcia2005} on time-series of $\langle \varepsilon^n\rangle_{xy,s}$ and $\langle \Omega^n\rangle_{xy,s}$, where $\langle \cdot \rangle_{xy,s}$ denotes a time- and $z$-dependent average over $x$-$y$ planes and symmetric halves of the channel ($s$). The block sizes used in the bootstrap analyses are approximately equal to one convective time unit ($L/U$) for both values of $Re_\tau$, giving ten snapshots per block for $Re_\tau\!=\!180$ and five snapshots per block for $Re_\tau\!=\!381$.} 

For both values of $Re_\tau$, Figure \ref{moms} shows that the $\varepsilon$ and $\Omega$ moments are large very close to the wall at $z^+\!=\!2$, but reach their minimum values at $z^+\!=\!10$ within the buffer layer. \joerg{The large amplitudes of the moments at $z^+\!=\!2$ arise from a combination of effects, including normalization of the moments by the mean values of $\varepsilon$ and $\Omega$, and the fact that $z^+\!=\!2$ corresponds to locations within the viscous sublayer where the flow is not fully turbulent. Moreover, while velocity fluctuations become small near the wall, fluctuating velocity gradients become very steep, resulting in very large values of $\varepsilon$ and $\Omega$ at $z^+\!=\!2$.} For both values of $Re_\tau$, the moments of $\varepsilon$ and $\Omega$ are similar up to $z^+\!\approx\! 30\!-\!40$, but for larger values of $z^+$, the moments of $\Omega$ are substantially greater than those for $\varepsilon$. This last result is consistent with the wider pdfs of $\Omega$ compared to those for $\varepsilon$ for $z^+\!\gtrsim\!40$ in Figure \ref{de}, and has also been observed in studies of homogeneous isotropic turbulence. \cite{donzis2008} 

While Figure \ref{moms} shows that the dependence of the moments on $z^+$ is similar in certain respects for $Re_\tau\!=\!180$ and $Re_\tau\!=\!381$, there are \joerg{slight} differences between the two values of $Re_\tau$. For $Re_\tau\!=\!381$, the moments of $\varepsilon$ reach a \joerg{local maximum} near $z^+\!\approx\! 50$ before decreasing until $z^+\!\approx\!100$. For larger values of $z^+$, the $\varepsilon$ moments remain relatively constant to the channel flow centerline. For $Re_\tau\!=\!180$, by contrast, the moments of $\varepsilon$ do not show a pronounced peak anywhere in the channel and remain relatively constant outside of the buffer layer. The local maxima in the $\varepsilon$ moments at $z^+\!\approx\! 50$ for $Re_\tau\!=\!381$ could indicate a qualitative change in the small-scale turbulence, in accordance with Yakhot {\it et al.}, \cite{yakhot2010} where it was suggested that small-scale fluctuations outside the buffer layer should be similar to those in isotropic turbulence. The invariance of the $\varepsilon$ and $\Omega$ moments for $z^+\!\gtrsim\! 100$ in the $Re_{\tau}\!=\!381$ simulation provides support for this idea. 

Figures \ref{moms}(a)-(c) further show that the moments of $\Omega$ for $Re_\tau\!=\!180$ reach a maximum near $z^+\!\approx\!110\!-\!150$ before decreasing again at the channel centerline. \joerg{This pronounced maximum is likely due to the bursting of coherent vortices, but is not observed for $Re_\tau\!=\!381$ in Figures \ref{moms}(d)-(f), where the moments of $\Omega$ remain relatively constant for $z^+\!\gtrsim\!100$. The error bars in Figure \ref{moms}(a)-(c) for $Re_\tau\!=\!180$ are, however, largest in this range of $z^+$, indicating that, particularly for the fourth moments, additional data is required to fully capture the strong temporal variability of the vorticity field created by these bursting structures.} 


\section{Conditional Analyses of Dissipation and Enstrophy Moments}
\label{struct}

The statistics of $\varepsilon$ and $\Omega$ in the previous section are obtained using all points in each plane of the channel. Certain features of these statistics, such as the local maxima in the moments of $\Omega$ for $Re_\tau\!=\!180$ at $z^+\!\approx\! 110\!-\!150$ and the differences between the moments for $\varepsilon$ and $\Omega$ outside of the buffer layer, may be due, in part, to the presence of intense vortical structures in the channel. This issue can be examined by calculating the statistical moments at points both within and outside regions of strong rotation in the flow, which are taken here to correspond to vortical structure locations.


\subsection{Identification of Intense Vortical Structures}
\label{id}
There has been considerable research over the last several decades on the most appropriate methods by which to identify intense vortical structures in turbulent flows. Chakraborty \textit{et al.} \cite{chakraborty2005} showed, however, that most vortical structure identification procedures give similar results for homogeneous isotropic turbulence. Since our primary interest here is not in characterizing the properties of the structures or assessing the merits of various identification procedures, we simply classify vortical structures as intense, rotation-dominated regions of the flow. The analysis of the $\varepsilon$ and $\Omega$ moments is then carried out only over points that fall within these regions, or points completely outside these regions, which we term the background flow.

As first proposed by Hunt \textit{et al.}, \cite{hunt1988} rotation-dominated regions can be identified as locations where $Q\!>\!0$, where $Q$ is the second invariant of the velocity gradient tensor, $A_{ij}\!=\! \partial u_i /\partial x_j$, and is written for an incompressible flow as
\begin{equation}
\label{qequation}
Q = \frac{1}{2}\left(-S_{ij}S_{ij} + R_{ij}R_{ij}\right)\,.
\end{equation}
The strain rate tensor, $S_{ij}$, is given by
\begin{equation}
\label{sijequation}
S_{ij} = \frac{1}{2}\left(\frac{\partial u_i}{\partial x_j} + \frac{\partial u_j}{\partial x_i}\right)\,,
\end{equation}
and $R_{ij}$ is the anti-symmetric part of $A_{ij}$ given by
\begin{equation}
\label{rijequation}
R_{ij} = \frac{1}{2}\left(\frac{\partial u_i}{\partial x_j} - \frac{\partial u_j}{\partial x_i}\right)\,.
\end{equation}
Here we use a closely related identification method, originally proposed by Zhou \textit{et al.}, \cite{zhou1999} that requires $A_{ij}$ to have two complex conjugate eigenvalues, $\lambda_r\!\pm\! i\lambda_{ci}$, and that the complex part, $\lambda_{ci}\!>\!0$, be larger than a prescribed cutoff. The restriction that $A_{ij}$ have two complex conjugate eigenvalues is equivalent to requiring that $\Delta\!>\!0$, where $\Delta$ is given by \cite{pirozzoli2008}
\begin{equation}
\label{dequation}
\Delta = Q^3+\frac{27}{4}R^2\,,
\end{equation}
and 
\begin{equation}
\label{requation}
R = -\frac{1}{3} \left(S_{ij}S_{jk}S_{ki} + 3R_{ij} R_{jk}S_{ki}\right)\,,
\end{equation}
is the third invariant of $A_{ij}$ for an incompressible flow. Regions of intense rotation -- and, by extension, vortical structure locations -- are then identified by requiring that $\lambda_{ci}\!>\!\alpha\left(\lambda_{ci}\right)_{max}$, where $\left(\lambda_{ci}\right)_{max}$ is the maximum value of $\lambda_{ci}$ at each value of $z^+$. The pre-factor $\alpha$ determines, in large part, the number and magnitude of vortical structures identified in the flow, \joerg{and we use $\alpha\!=\!0.05$ herein.} The resulting locations of intense rotation, identified as points where $\lambda_{ci}$ is large, are generally similar to those obtained using other criteria, including the $Q$ criterion. \cite{chakraborty2005}

Due to the presence of a mean shear in the channel, the entire vortical structure identification procedure is carried out using the fluctuating velocity gradient, $A'_{ij}\!\equiv\!\partial u'_i /\partial x_j$. This approach has been used previously by Robinson \cite{robinson1991} and Pirozzoli \textit{et al.}, \cite{pirozzoli2008} and is particularly important near the channel walls where the mean shear can generate large vorticity not associated with coherent vortical structures. The fluctuating tensors $S'_{ij}$, given in Eq.\ (\ref{sijf}), and $R'_{ij}$, given by
\begin{equation}
\label{rijf}
R'_{ij} = \frac{1}{2}\left(\frac{\partial u'_i}{\partial x_j} - \frac{\partial u'_j}{\partial x_i}\right)\,,
\end{equation}
are thus used in the expressions for $Q$, $R$, and $\Delta$ in Eqs.\ (\ref{qequation}), (\ref{requation}), and (\ref{dequation}). The difference between using the full velocity gradient tensor and the fluctuating tensor is small for much of the channel, and only becomes significant in the near-wall region where the mean shear is large.

\subsection{Orientation of Intense Vortical Structures}

\begin{figure}[t]
\begin{center}
\includegraphics[width=10cm]{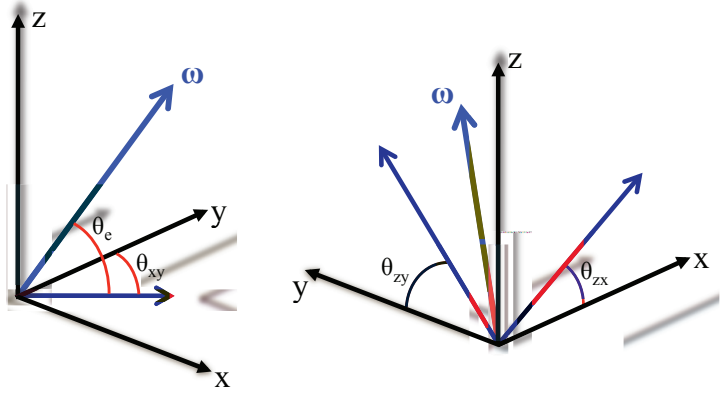}
\caption{Schematic showing vorticity orientation angles defined in Eq.\ (\ref{anglesdef}).}
\label{angles}
\end{center}
\end{figure}

In the present analysis, the orientation of intense vortical structures is inferred from the orientation of the fluctuating vorticity field, $\omega'_i$, at rotation-dominated locations in the flow. This is in contrast to prior approaches (e.g., Ref.\ \cite{ganapathi2006}) which have attempted to treat coherent vortices as connected regions with a single orientation associated with the structure as a whole. Following Pirozzoli \textit{et al.}, \cite{pirozzoli2008} the orientation of the vorticity is characterized using the angles 
\begin{equation}\label{anglesdef}
\theta_{ij} = \tan^{-1}\left(\frac{\omega'_i}{\omega'_j}\right)\,,\quad
\theta_{e} = \sin^{-1}\left(\frac{\omega'_z}{\omega'}\right)\,,
\end{equation}
where $\omega'\equiv \left(\omega'_i \omega'_i\right)^{1/2}$, which are shown schematically in Figure \ref{angles}. The orientation of intense vortical structures can then be determined from the joint pdfs of $\theta_{xy}$ and $\theta_e$, denoted $P(\theta_{xy},\theta_{e})$. Since an isotropic vorticity field gives $P(\theta_{xy},\theta_{e})\!\sim\!\cos(\theta_e)$, we consider normalized joint pdfs of $P(\theta_{xy},\theta_{e})/\cos(\theta_e)$, following the approach used in Ref.\ \cite{pirozzoli2008}. 

\begin{figure}[t]
\begin{center}
\includegraphics[]{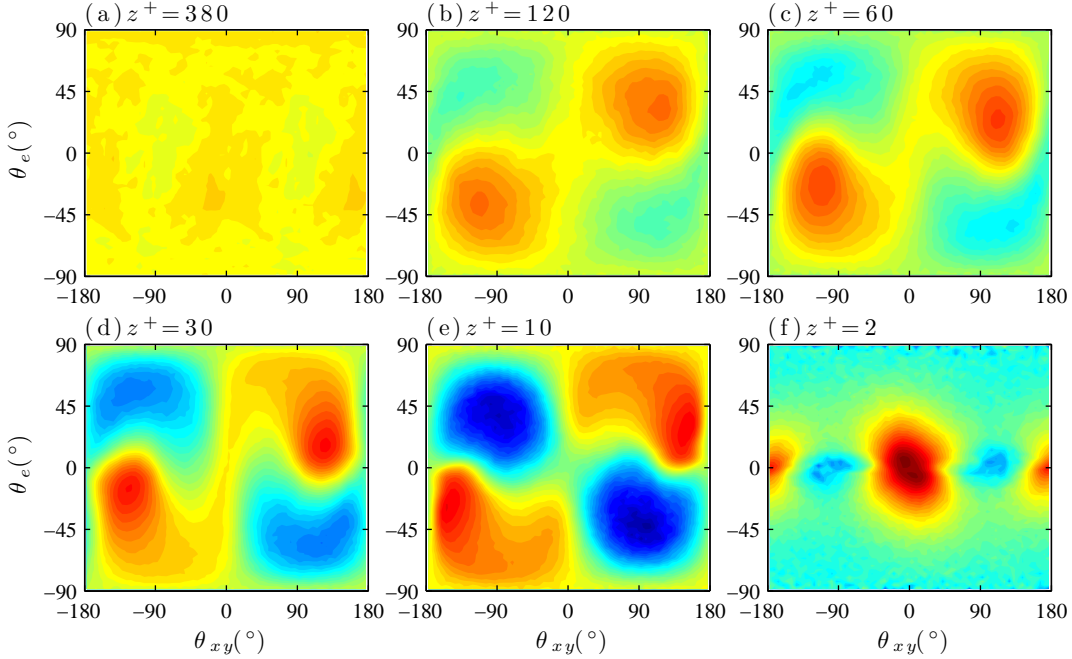}
\caption{Joint pdfs of the elevation angle $\theta_e$ and the orientation angle $\theta_{xy}$, defined in Eq.\ (\ref{anglesdef}), at six values of $z^+$ for $Re_\tau=381$. Only vorticity identified to be within vortical structures is included in the pdfs (i.e.\ points for which $\lambda_{ci}>0.05 (\lambda_{ci})_{max}$, where $(\lambda_{ci})_{max}$ is determined at each $z^+$). Contours of $\log_{10}\left[P(\theta_{xy},\theta_e)/\cos\theta_e\right]$ are plotted, with levels from $10^{-3}$ (blue) to $10^{0}$ (red).} \label{vort_struct_orient}
\end{center}
\end{figure}

Joint pdfs of $P(\theta_{xy},\theta_{e})/\cos(\theta_e)$ for $Re_\tau\!=\!381$ in Figure \ref{vort_struct_orient} show that the distribution of vorticity within rotation-dominated regions varies substantially as the wall is approached from the channel centerline. There is essentially no preferred orientation of the vorticity near the centerline, as shown by the lack of any clear maxima or minima in Figure \ref{vort_struct_orient}(a). At $z^+\!\approx\! 120$ in Figure \ref{vort_struct_orient}(b), however, the vorticity shows a weak preference for $\theta_e\!\approx\! \pm 45^\circ$ and $\theta_{xy}\!\approx\! \pm 90^\circ$, approximately corresponding to the ``necks'' of hairpin vortices discussed in previous studies. \cite{ganapathi2006,pirozzoli2008} \joerg{As the wall is approached, Figure \ref{vort_struct_orient}(c) shows that the peaks in the distributions begin to shift towards $\theta_e\!=\!0^\circ$, corresponding to quasi-streamwise vortices \cite{pope2000,pirozzoli2008} which are oriented in the $x$-direction parallel to the channel walls. At the same time, the probability for $\theta_{xy}\!>\!|90^\circ|$ also begins to increase, and by $z^+\!=\!10$ in Figure \ref{vort_struct_orient}(e) there is a strong probability of obtaining vorticity with $\theta_{xy}$ close to $\pm180^\circ$. At $z^+\!=\!2$ in Figure \ref{vort_struct_orient}(f), much of the vorticity is oriented in the spanwise direction parallel to the channel walls (giving $\theta_e\!\approx\!\theta_{xy}\! \approx\! 0^\circ$). There are also less pronounced peaks at $|\theta_{xy}|\!=\!180^\circ$, which are due to weaker vorticity not associated with intense structures (as indicated by the disappearance of these secondary peaks when using larger values of $\alpha$). There is thus a rapid change in the orientation of intense vortices between the buffer layer and viscous sublayer, a result that is mirrored in the large changes in the moments of $\varepsilon$ and $\Omega$ between $z^+\!=\!10$ and $z^+\!=\!2$ in Figure \ref{moms}.}

\begin{figure}[t]
\begin{center}
\includegraphics[]{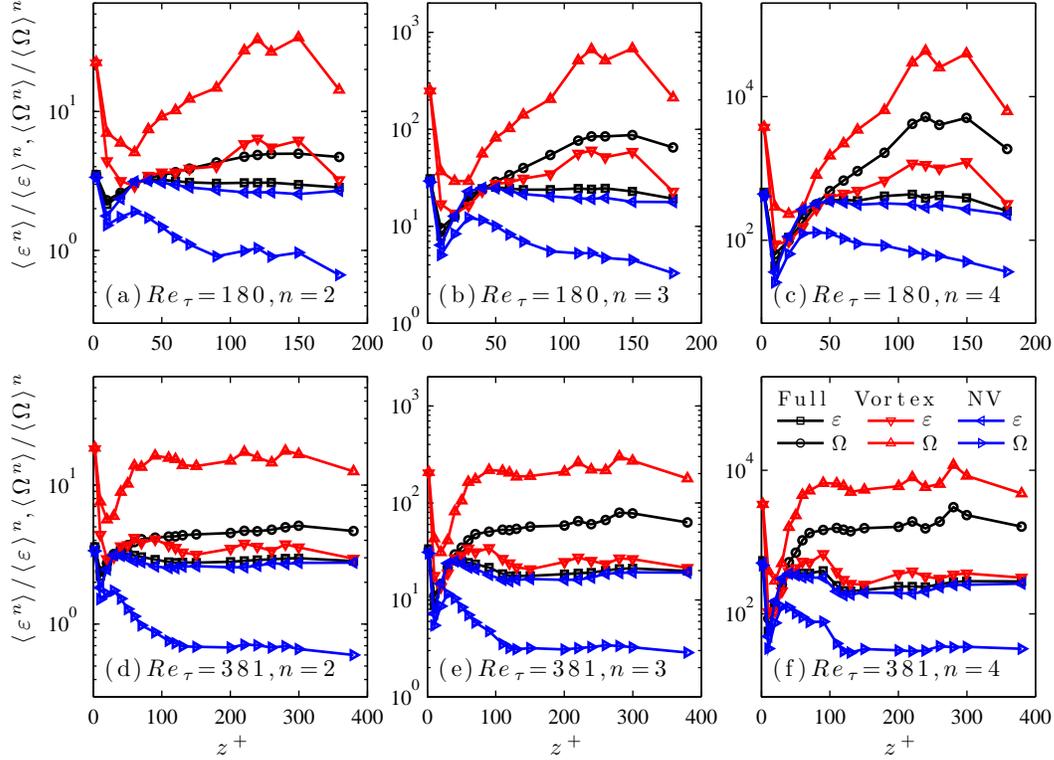}
\caption{Moments $\langle \varepsilon^n\rangle /\langle \varepsilon\rangle^n$ and $\langle \Omega^n\rangle /\langle \Omega\rangle^n$ for $n=2-4$ in the $Re_\tau=180$ (a)-(c) and $Re_\tau=381$ (d)-(f) simulations. The curves correspond to moments obtained from the full fields (black lines, denoted ``Full''), points at rotation-dominated locations (red lines, denoted ``Vortex''), and points where the rotation is weak (blue lines, denoted ``NV'').}\label{moms_struct}
\end{center}
\end{figure}

\subsection{Conditioned Moments of the Dissipation and Enstrophy}
\label{cond}

Figure \ref{moms_struct} shows the moments of $\varepsilon$ and $\Omega$ conditioned on vortical structure locations, which are identified as rotation-dominated points in the flow using the procedure outlined in Section \ref{id}. Moments are also shown for points in the background field, where the rotation is small (see the $\lambda_{ci}$ criterion defined immediately below Eq.\ (\ref{requation})). The moments for both subsets are normalized with respect to the means of $\varepsilon$ and $\Omega$ obtained from the full field. 

For both $\varepsilon$ and $\Omega$, Figure \ref{moms_struct} shows that the moments obtained at locations of intense rotation exceed those for both the full and background fields. Although the increase in the $\Omega$ moments at these locations is to be expected (since the identification procedure essentially selects points with large vorticity magnitude), the accompanying increase in $\varepsilon$ can be understood from the joint pdfs in Figure \ref{de_joint}. These pdfs show that intense $\varepsilon$ and $\Omega$ are statistically correlated, and thus the moments of $\varepsilon$ at locations of intense rotation tend to be larger than the corresponding full field values. From studies of isotropic turbulence (see, e.g., \cite{schumacher2010}), it is also well-known that large amplitudes of both $\varepsilon$ and $\Omega$ appear in close spatial proximity. At the same time, however, it should be noted that the moments of $\varepsilon$ remain smaller than those of $\Omega$ at all $z^+$, for all orders, and at both values of $Re_\tau$. 

Substantial differences between the moments of $\Omega$ and $\varepsilon$ are observed in the background field, where the rotation is small. In particular, there is near agreement between the $\varepsilon$ moments in the full and background fields. This indicates that locations of intense rotation, and hence vortical structures, do not contribute substantially to the full statistics of $\varepsilon$. By contrast, Figure \ref{moms_struct} shows that the background moments of $\Omega$ are significantly smaller than the corresponding full field values, indicating that locations within vortical structures make a substantial contribution to the full field moments of $\Omega$. \joerg{Contrary to the moments in the full field and at rotation dominated locations, the local maxima in the moments of $\Omega$ at $z^+\!\approx\! 150$ for $Re_\tau\!=\!180$ are absent in the background field. This reinforces the connection between these maxima and intense vortices in the flow. Such pronounced maxima are not observed in any of the fields shown in Figure \ref{moms_struct} for $Re_\tau\!=\!381$, and it remains to be seen in future studies whether these results undergo additional changes for even larger $Re_{\tau}$.}

\section{Summary and Conclusions}
\label{conc}
The present high-resolution DNS study of fully-developed turbulent channel flow has examined velocity gradient statistics as a function of wall distance, $z^+$, at friction Reynolds numbers $Re_\tau\!=\!180$ and $Re_\tau\!=\!381$. An emphasis has been placed on the statistics, and in particular the higher-order moments, of the energy dissipation rate, $\varepsilon$, and the local enstrophy, $\Omega$. \joerg{The probability density functions (pdfs) of $\varepsilon$ and $\Omega$ for the $Re_\tau\!=\!381$ case are qualitatively similar to previous results obtained for $Re_\tau\!=\!180$. \cite{boeck2010} The pdfs of both $\varepsilon$ and $\Omega$ generally become less intermittent as the wall is approached from the channel centerline, as indicated by the less broad tails of the pdfs. For $z^+\!\gtrsim\!40$, the pdfs of $\Omega$ have wider tails than the pdfs of $\varepsilon$, although the pdfs for both quantities are similar near the channel walls.} Joint pdfs show that both large and small values of $\varepsilon$ and $\Omega$ are correlated, and that the support of the pdfs is reduced as the wall is approached.

The high resolution of the present simulations has allowed moments of $\varepsilon$ and $\Omega$ up to fourth order to be calculated. The moments of $\varepsilon$ and $\Omega$ are similar in the viscous sublayer and buffer layer, but, for locations further from the wall, the moments of $\Omega$ are substantially larger than those of $\varepsilon$. Reynolds number effects are observed in the moments of both $\varepsilon$ and $\Omega$. In particular, there is a local maximum in the moments of $\varepsilon$ \joerg{at the beginning} of the logarithmic layer for $Re_\tau\!=\!381$, which is not present in the moments of $\varepsilon$ for $Re_\tau\!=\!180$. For both values of $Re_\tau$, the moments of $\varepsilon$ remain relatively constant from $z^+\approx 100$ to the channel centerline. The $\Omega$ moments also remain relatively constant over this range for $Re_\tau\!=\!381$, but for $Re_\tau\!=\!180$ the moments of $\Omega$ increase and reach a maximum at $z^+\!\approx\! 110\!-\!150$, before decreasing at the channel centerline. \joerg{The maxima in the moments of $\Omega$ for the smallest $Re_\tau$ are most likely due to bursting of vortical structures across the channel, which is an effect due to the low Reynolds number. In particular, it is possible that both wall regions may not be fully decoupled. This bursting causes large temporal variations of the local enstrophy, which complicates the statistical convergence of higher order moments. This lack of convergence may thus have a physical fingerprint, and substantially more temporal snapshots (requiring significant additional computational effort) are necessary to resolve this issue in future investigations.}

Using conditional analyses based on regions of intense rotation, which are taken here to correspond to vortical structure locations, the moments of both $\varepsilon$ and $\Omega$ are shown to be larger in rotation-dominated regions than in the full field. The increase in $\varepsilon$, in particular, is due to the correlation between events of intense $\varepsilon$ and $\Omega$ shown in Figure \ref{de_joint}. At the same time, moments calculated in the background field, where the rotation is not intense, indicate that vortical structures make only a small contribution to the moments of $\varepsilon$ in the full field. \joerg{Differences between the moments for the two values of $Re_\tau$ in the conditional intense rotation fields are similar to the differences in the full field. In particular, the conditional moments of $\varepsilon$ and $\Omega$ are relatively constant for $z^+\!\gtrsim\! 100$ in the $Re_\tau\!=\!381$ case, but local maxima, particularly in the moments of $\Omega$, are still observed at $z^+\!\approx\! 150$ for the $Re_\tau\!=\!180$ case. These maxima are not present, however, in the $\Omega$ moments in the background field for $Re_\tau\!=\!180$.}

The next step in the analysis is to extend the study of the Reynolds number dependence of the statistics, which requires further high-resolution data at larger values of $Re_{\tau}$. This will allow us to make comparisons with similar studies in isotropic turbulence, \cite{schumacher2007} and to identify quantitative differences in the statistics close to the channel walls. These simulations, which are currently in progress, require substantial computational resources and will be reported in the near future.
\\
\\
{\em Acknowledgements} \\

We thank the DEISA Consortium (www.deisa.eu), co-funded through the EU FP6 project RI-031513 and the FP7 project RI-222919, for support within the DEISA Extreme Computing Initiative. The computations were carried out on the Cray XT4 Cluster Hector at EPCC in Edinburgh. We wish to thank Florian Janetzko for assistance with the parallel input/output routines. PEH was supported by the National Research Council Research Associate Program of the National Academy of Sciences. DK and TB were supported by the Emmy-Noether-Program and JS was supported by the Heisenberg-Program of the Deutsche Forschungsgemeinschaft (DFG).

\bibliographystyle{unsrt}

\end{document}